\documentclass{optica-article}

\journal{opticajournal} 

\articletype{Research Article}
\providecommand{\authormark}[1]{$^{#1}$}

\begin{document}

\title{Real-Time Dynamic Crosstalk Tracking and Compensation via Photonic Blind Source Separation}

\author{Dongliang Wang\authormark{1,$\dag$}, Yikun Nie\authormark{1,$\dag$}, Dingqian Zhao\authormark{1,$\dag$}, Dan Yi\authormark{1}, Gaolei Hu\authormark{1},  Hon Ki Tsang\authormark{1}, and Chaoran Huang\authormark{1,*}}

\address{\authormark{1}Department of Electronic Engineering, The Chinese University of Hong Kong, Shatin, Hong Kong.\\
\authormark{$\dag$}The authors contributed equally to this work.}

\email{\authormark{*}crhuang@ee.cuhk.edu.hk} 


\begin{abstract*} 
Space-division multiplexing in few-mode fibers can substantially increase optical-link capacity, but its practical deployment is hindered by dynamically varying inter-modal crosstalk induced by environmental perturbations. Although silicon photonic reconfigurable mode processors have been widely demonstrated, continuous adaptation during high-speed intensity-modulation direct-detection (IM/DD) transmission remains challenging. Here, we experimentally demonstrate real-time tracking and compensation of dynamically varying inter-modal crosstalk using a hybrid photonic--electronic framework that combines an integrated silicon photonic processor with an FPGA-based control backend. The photonic processor performs signal separation in the optical domain, while the FPGA extracts signal statistics and implements a blind source separation (BSS)-based feedback algorithm without requiring dedicated training sequences. A two-stage optimization strategy combines rapid crosstalk suppression with stable continuous tracking under dynamic channel variations. We demonstrate photonic blind source separation for signals up to 100~GBaud and FPGA-enabled continuous adaptation for two 64-GBaud data channels in a few-mode-fiber transmission system. The system maintains bit-error rates below $10^{-4}$ under dynamic crosstalk, with a per-update control latency of 4~ms. These results establish a practical route toward low-latency, hardware-efficient adaptive photonic front ends for dynamic high-speed IM/DD space-division-multiplexed links.

\end{abstract*}

\section{Introduction}

The rapid evolution of artificial intelligence (AI) has driven a substantial increase in the demand for high-capacity data transmission ~\cite{kachris2012survey, shekhar2024roadmapping}. Optical interconnects have thus become indispensable, offering significantly higher bandwidth density and lower transmission loss over long distances compared with electrical links ~\cite{zhao2024development}. To further scale the capacity of fiber-optic communication systems, space-division multiplexing (SDM) has emerged as a promising paradigm ~\cite{richardson2013space,willner2021perspectives}.  By exploiting the spatial dimension of optical fibers, SDM enables multiple orthogonal spatial channels to carry independent data streams in parallel, thereby overcoming the capacity limits of conventional single-mode fibers.

\begin{figure*}[htbp]
\centering
  \includegraphics[width=1\linewidth]{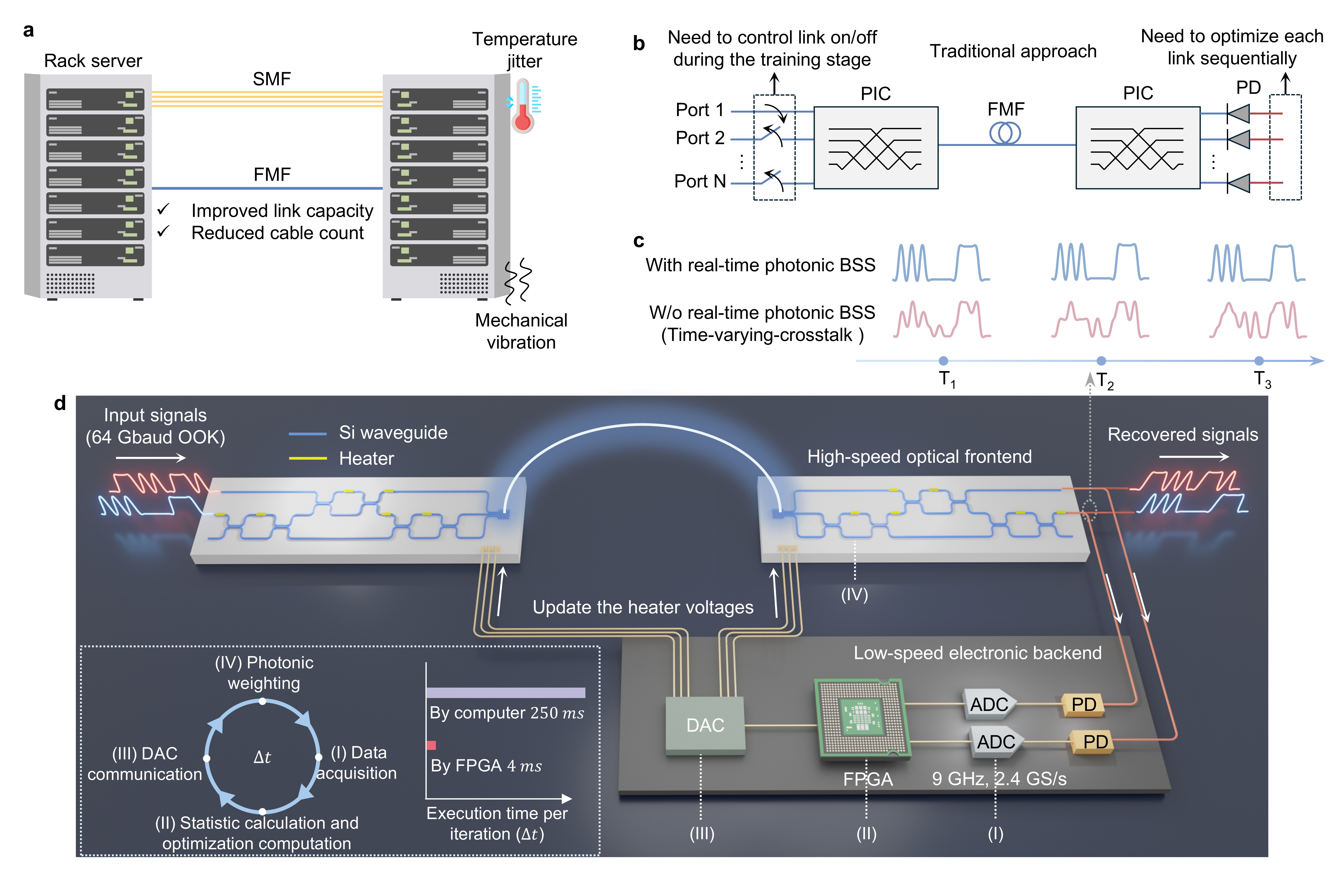}
  \caption{\textbf{Schematic illustrations of the proposed hybrid photonic–electronic system enabling real-time dynamic crosstalk tracking and compensation.}
\textbf{a}, Illustration of a few-mode-fiber link. Few-mode-fiber links enable space-division multiplexing with increased link capacity and reduced cable count. Environmental perturbations, such as temperature variation and mechanical vibration, induce time-varying inter-modal crosstalk. \textbf{b}, Traditional mode demultiplexing approach. Conventional photonic demultiplexing requires sequential activation of the input channels during training and separate optimization of each output channel at the receiver. Such link switching disrupts the data stream and is therefore unsuitable for real-time communication. \textbf{c}, Comparison of signal recovery with and without real-time photonic BSS under time-varying crosstalk. With real-time adaptation, the recovered signals remain stable at different time instants. Without real-time update, the output signals are distorted and vary with the crosstalk condition. \textbf{d}, Proposed hybrid photonic–electronic system. An integrated silicon-photonic MZI mesh performs optical-domain mode demultiplexing. A low-speed electronic backend with PDs, ADCs, DACs, and an FPGA computes signal statistics and updates the mesh heater voltages in real time. The FPGA reduces the per-iteration control latency from about 250 ms to 4 ms.
}
  \label{figure 1}
\end{figure*}

However, SDM systems inherently suffer from inter-channel crosstalk caused by fiber structural imperfections and external disturbances ~\cite{matthes2021learning}. More importantly, the crosstalk fluctuates randomly over time due to environmental perturbations such as temperature fluctuations and mechanical vibrations ~\cite{carpenter2015observation, ryf2012mode}, as shown in Fig.~\ref{figure 1}a. As a result, the optical transmission channel becomes time-varying, posing significant challenges for reliable optical communication systems. To address this issue, digital signal processing (DSP) is traditionally employed, where multiple-input multiple-output (MIMO) algorithms are used to mitigate inter-channel crosstalk and separate the mixed signals ~\cite{ryf2012mode, randel20116}. However, the high computational complexity of DSP-based MIMO processing leads to significant power consumption and latency, posing a major bottleneck for real-time operation in next-generation optical systems.

Recent advances in photonic integrated circuits (PICs) offer a promising solution to these limitations ~\cite{shastri2021photonics,huang2021silicon,xu202111,huang2022prospects,zhou2022photonic, chen2023all, wang2024ultrafast}. Photonic processors support significantly higher bandwidth, improved energy efficiency, and ultra-low-latency signal processing ~\cite{miller2009device}. In SDM systems, on-chip Mach–Zehnder interferometer (MZI) meshes have been employed to mitigate mode mixing and separate coupled spatial channels ~\cite{miller2013self,wu2023chip,seyedinnavadeh2024determining,lavery2024re,lu2024empowering,wan2024efficient,ruan2024flexible, huang2022high, wang2022photonic}. However, many existing implementations are limited to compensating static crosstalk, as they rely on pre-calibration or offline training and therefore cannot track and compensate for dynamic crosstalk in SDM. As illustrated in Fig.~\ref{figure 1}b, the conventional training procedure isolates individual channels by turning on the transmitters one at a time, followed by separate optimization of each channel path. This process disrupts continuous data transmission, rendering real-time signal processing impractical. More recently, dynamic all-optical MIMO demultiplexing has been demonstrated for coherent Quadrature Phase Shift Keying (QPSK) transmission over randomly coupled multi-core fiber, showing the potential of adaptive photonic processors for SDM systems~\cite{grillanda2025transmission}. Nevertheless, such approaches rely on interference-induced optical power variations as feedback signals for optimization, which are mainly applicable to coherent transmission systems. In IM/DD links, the optical intensity itself carries the data information, making it difficult to distinguish crosstalk-induced intensity fluctuations from data modulation using direct power-based feedback.

Achieving real-time adaptation of photonic processors during continuous IM/DD signal transmission requires overcoming both system-level and algorithmic challenges. At the system level, conventional oscilloscope-based acquisition with external computer processing introduces substantial latency because large volumes of signal data must be buffered, stored, and transferred before optimization. At the algorithmic level, robust online optimization strategies capable of tracking rapidly varying SDM crosstalk remain limited. This challenge becomes more severe under strong mode coupling, where fiber-induced phase perturbations between coupled optical fields are converted into temporal intensity fluctuations through optical interference, thereby destabilizing the optimization process. Together, these limitations prevent existing photonic processors from effectively addressing dynamically varying crosstalk in high-speed IM/DD SDM transmission systems.

Here, we experimentally demonstrate a hybrid photonic--electronic system that enables real-time crosstalk tracking and mitigation for high-speed IM/DD SDM transmission. The system integrates an on-chip silicon photonic processor with an FPGA-based controller implementing a photonic blind source separation (BSS) algorithm, as illustrated in Fig.~\ref{figure 1}d. The on-chip MZI mesh performs the required linear transformation in the optical domain, while the FPGA adaptively updates the photonic weights to compensate for dynamically varying crosstalk. Leveraging the parallel processing capability of the FPGA, the control loop is implemented as a memory-less streaming architecture, where the signal statistics required for optimization are extracted directly from the ADC data stream on-the-fly. This approach eliminates the memory buffering and data-transfer overhead associated with conventional oscilloscope-based acquisition and software-based processing.

The BSS algorithm separates independent source signals from observed mixtures by exploiting their statistical independence, without requiring prior knowledge of the mixing process or dedicated training sequences~\cite{choi2005blind,zhang2023broadband,lederman2023real,zhang2024system}. Since the optimization relies on signal statistics rather than full waveform reconstruction, the electronic backend requirements in sampling rate and bandwidth are substantially relaxed, enabling a medium-speed FPGA to support signals beyond 64~GBaud. To ensure stable optimization under dynamic fiber perturbations, we propose a two-stage optimization strategy. A derivative-free global search first rapidly suppresses strong crosstalk, followed by gradient-based updates for efficient continuous tracking once the system approaches a low-crosstalk operating point. This strategy enables stable adaptation without interrupting ongoing data transmission.

Experimentally, we demonstrate dynamic crosstalk mitigation for 100~GBaud OOK signals in a few-mode-fiber transmission system. We further realize FPGA-enabled real-time adaptation for 64~GBaud signals under dynamic crosstalk conditions, maintaining bit-error rates below $10^{-4}$ throughout the observation period. Compared with conventional software-based control schemes, the FPGA streaming architecture eliminates external data transfer and memory buffering, reducing the adaptation latency by more than 60-fold and enabling millisecond-scale feedback updates. Moreover, unlike digital equalizers whose hardware complexity scales with signal bandwidth~\cite{ryf2012mode, randel20116}, the latency and energy consumption of the proposed control architecture are largely independent of the channel data rate. These features establish a scalable approach for real-time crosstalk tracking and compensation in next-generation high-capacity SDM optical links.

\section{Principle}

Mode-division multiplexing (MDM) is a promising implementation of SDM, where independent data streams are encoded onto the orthogonal transverse eigenmodes of few-mode fibers (FMFs) to increase transmission capacity. In practical FMF links, however, fiber imperfections and environmental perturbations, such as thermal fluctuations and mechanical vibrations, induce coupling among spatial modes and break their orthogonality. As a result, transmitted channels become mixed during propagation, leading to dynamically varying inter-modal crosstalk. Together with mode-dependent loss (MDL), the ideal diagonal transmission matrix of an FMF link becomes a dense, time-varying, and generally non-unitary mixing matrix. The input--output relationship of the spatial channels can therefore be expressed as

\begin{equation}
\mathbf{y}(t)=\mathbf{H}(t)\mathbf{x}(t),
\end{equation}

where $\mathbf{x}(t)$ and $\mathbf{y}(t)$ denote the transmitted and received signal vectors, respectively. Since $\mathbf{H}(t)$ continuously evolves with the fiber state, static compensation schemes cannot maintain reliable channel separation, making real-time tracking and compensation of the channel transformation essential.

A general approach to recover the transmitted channels is to decompose the channel matrix through singular value decomposition (SVD),

\begin{equation}
\mathbf{H}(t)=\mathbf{U}(t)\mathbf{\Sigma}(t)\mathbf{V}^{\dagger}(t),
\end{equation}

where $\mathbf{\Sigma}(t)$ represents the transmission coefficients of independent eigenchannels, while $\mathbf{U}(t)$ and $\mathbf{V}(t)$ define the output and input eigenchannel transformations, respectively. By implementing the corresponding inverse transformations, the coupled spatial channels can be transformed into independent transmission paths:

\begin{equation}
\mathbf{U}^{\dagger}(t)\mathbf{H}(t)\mathbf{V}(t)=\mathbf{\Sigma}(t).
\end{equation}

Reconfigurable photonic processors provide an efficient hardware platform for implementing these adaptive transformations. In particular, MZI meshes can realize programmable unitary transformations and have therefore been widely explored for mode manipulation, channel separation, and crosstalk suppression in SDM systems. However, implementing programmable photonic processors alone is insufficient for dynamic SDM links; the mesh configurations must be continuously adapted to track the time-varying channel transformation.

In principle, this adaptation requires continuous knowledge of the instantaneous channel matrix $\mathbf{H}(t)$. However, obtaining explicit channel state information in rapidly varying FMF links is challenging due to the required channel estimation overhead, processing latency, and feedback bandwidth. To overcome this limitation, we employ a blind source separation (BSS) algorithm that directly optimizes the photonic processor based on received signal statistics without requiring explicit channel estimation. The underlying principle originates from the Central Limit Theorem: independent transmitted data streams are typically non-Gaussian, whereas linear mixing tends to increase the Gaussianity of the received signals. Therefore, by maximizing a non-Gaussianity metric, such as kurtosis, the photonic processor can be iteratively optimized toward a demixing transformation.

Importantly, because BSS optimization relies on statistical features rather than full waveform information, the feedback metrics can be extracted from low-speed and sparsely sampled measurements. This substantially relaxes the bandwidth and sampling-rate requirements of the electronic backend. Furthermore, since these statistical features evolve according to the channel variation rather than the symbol rate, the feedback update rate is determined by the timescale of channel dynamics instead of the transmitted baud rate. The required statistics can therefore be computed directly from the ADC data stream using an FPGA-based streaming architecture, avoiding waveform reconstruction, memory buffering, and inter-device data transfer. This enables low-latency adaptation of the photonic processor for real-time dynamic crosstalk compensation in high-speed IM/DD SDM transmission systems.

 \section{Experimental setup}

\begin{figure*}[htbp]
\centering
  \includegraphics[width=1\linewidth]{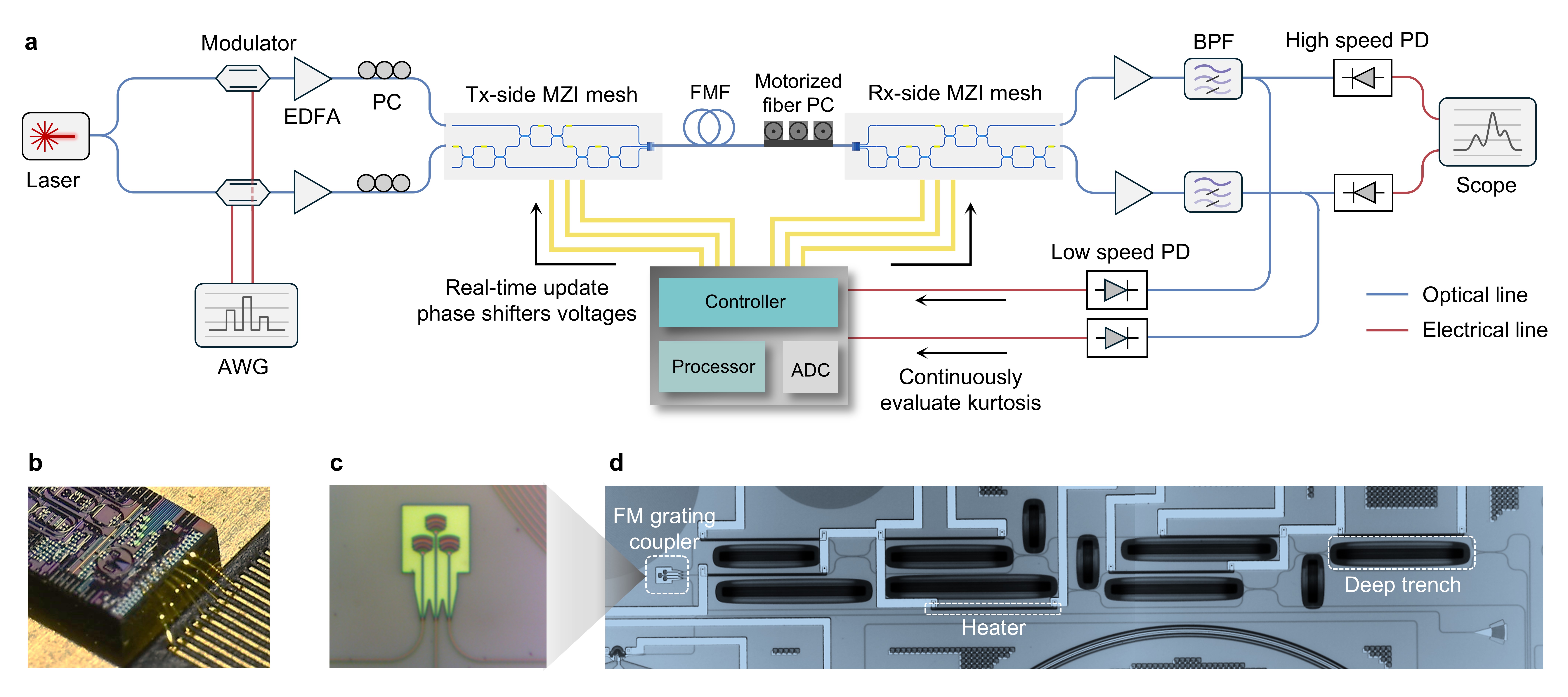}
  \caption{\textbf{Device design and experimental setup.}
\textbf{a}, Schematic of the experimental setup. Two modulated optical channels are injected into a silicon-photonic MZI mesh for mode demultiplexing. The demultiplexed outputs are monitored by a low-speed electronic feedback loop, which continuously evaluates kurtosis and updates the phase-shifter voltages in real time to track time-varying crosstalk. High-speed photodiodes and an oscilloscope are used for waveform characterization. \textbf{b}, Photograph of the packaged silicon-photonic chip. \textbf{c}, Microscope image of the focusing FM grating coupler. \textbf{d}, Microscope image of the photonic circuit, highlighting the few-mode grating coupler, heater, and deep-trench structures.
}
  \label{figure 2}
\end{figure*}

The experimental setup is illustrated in Fig.~\ref{figure 2}a. The system consists of a high-speed optical transmitter, programmable silicon photonic processors at the transmitter (Tx) and receiver (Rx), a FMF link with emulated dynamic perturbations, and a low-speed electronic feedback loop. In this architecture, high-bandwidth signal processing is carried out entirely in the analog optical domain, while adaptive optimization is performed by the electronic controller.

At the transmitter, a continuous-wave (CW) laser is split into two paths. Each path passes through a polarization controller (PC) and a thin-film lithium niobate (TFLN) Mach-Zehnder modulator (MZM; Liobate LB4C6PSBM63, 40~GHz bandwidth, RF \(V_\pi \approx 3\)~V). Driven by an arbitrary waveform generator (AWG; Keysight M8199A, 256~GSa/s), the two MZMs generate OOK signals up to 100~GBaud. The modulated signals are then amplified by erbium-doped fiber amplifiers (EDFAs), polarization-adjusted by additional PCs, and injected into the Tx-side photonic processor.

The Tx and Rx processors are implemented as programmable silicon photonic MZI meshes with integrated thermo-optic phase shifters for independent phase control. Detailed information on the silicon photonic processor is provided in Appendix~A. At the Tx side, the mesh transforms the input signals into spatial modes for transmission through a 5-m graded-index few-mode fiber (FMF) link. The FMF supports three spatial modes, including the fundamental LP$_{01}$ mode and the nearly degenerate LP$_{11}$ mode group consisting of LP$_{11a}$ and LP$_{11b}$. A nanoantenna-array-based few-mode grating coupler is designed to interface the silicon photonic processors with the FMF by enabling optical coupling between the on-chip waveguides and the spatial modes supported by the FMF. Detailed specifications of the FMF and grating coupler are provided in Appendix~B and~C, respectively. To emulate time-varying channel perturbations in a stable short-reach link, a motorized polarization controller (MPC320, Thorlabs) is inserted before the Rx-side processor to introduce controlled dynamic mode coupling and modal crosstalk.

After propagation through the FMF, the received optical signals are processed by the Rx-side MZI mesh and then coupled out through single-mode grating couplers. The demultiplexed outputs are amplified by EDFAs, filtered by optical bandpass filters (BPFs), and detected by high-speed photodiodes (Coherent XPDV3120R-VM-FA, 70~GHz bandwidth). The electrical waveforms are subsequently captured by a real-time oscilloscope (Keysight UXR0592AP, 59~GHz bandwidth, 256~GSa/s).

Adaptive control of the photonic meshes is implemented through the electronic feedback loop shown in the lower part of Fig.~\ref{figure 2}a. At the output of the Rx processor, 30\% of the optical power is tapped and detected by low-speed photodiodes (10~GHz bandwidth). The photocurrents are digitized by a 9~GHz ADC operating at 2.4~GSa/s and streamed to an FPGA for real-time statistical processing. Based on the extracted signal statistics, the control algorithm determines the required phase updates, which are converted by a multi-channel digital-to-analog converter (DAC) into drive voltages for the thermo-optic phase shifters on both the Tx and Rx meshes.

The photonic processors are electrically bonded for operation, as shown in Fig.~\ref{figure 2}b. Optical coupling is achieved through standard single-mode grating couplers at the single-mode ports and a focusing FM grating coupler at the FMF interface. A microscope image of the focusing FM grating coupler is shown in Fig.~\ref{figure 2}c, and the micrographs in Fig.~\ref{figure 2}d highlight key on-chip components, including the FM grating coupler, thermo-optic heaters, and deep-trench thermal isolation structures.

\section{Two-Stage Optimization Framework: real-time signal compensation without disrupting transmission}

\begin{figure*}[htbp]
\centering
  \includegraphics[width=1\linewidth]{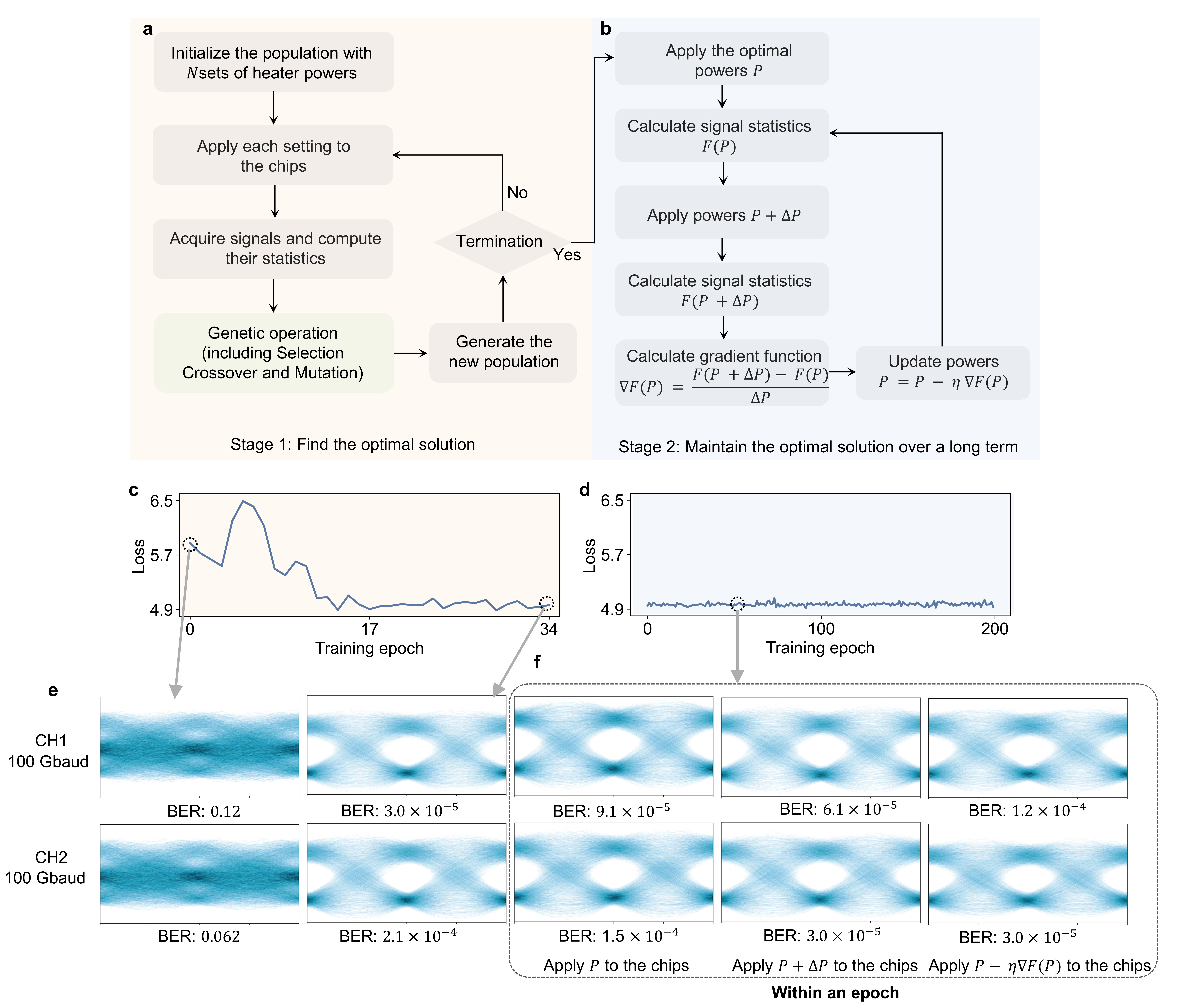}
  \caption{\textbf{Two-stage optimization framework.}
\textbf{a}, Stage I flowchart for searching the optimal heater voltages using a genetic algorithm. \textbf{b}, Stage II flowchart for maintaining the optimal solution through gradient-based voltage updates. \textbf{c},\textbf{d}, Loss evolution during stage I and stage II, respectively. \textbf{e}, Eye diagrams of the two channels before and after stage I optimization. \textbf{f}, Eye diagrams of the two channels within one update cycle in stage II, corresponding to the application of \(P\), \(P + \Delta P\), and \(P - \eta \nabla F(P)\), respectively.
}
  \label{figure 3}
\end{figure*}

In practical optical networks carrying real-time traffic, optimization must continuously track environmental drifts without interrupting data transmission. Gradient-based methods are well suited to this scenario because they can update the control parameters in real time using directly measured feedback signals. In the gradient descent (GD) algorithm, the gradient of the cost function \(F\) is typically estimated by finite difference:
\begin{equation}
    \nabla F(P) = \frac{F(P + \Delta P) - F(P)}{\Delta P},
\end{equation}
where, in our experimental setting, \(P\) is the heater-power vector, and \(\Delta P\) is a small perturbation applied to it. This method assumes that the system state remains unchanged between the measurements of \(F(P)\) and \(F(P + \Delta P)\) except for the applied perturbation.

However, implementing real-time optimization in an interferometric optical system faces an additional challenge arising from phase fluctuations in the multi-path optical link. The relative phases between different optical paths at the Tx side can drift over time due to environmental perturbations, such as thermal fluctuations and mechanical vibrations. Under high-crosstalk conditions, interference between the coupled optical fields converts these phase fluctuations into large intensity variations at the detector. As a result, the cost function values $F(P)$ and $F(P+\Delta P)$ may be evaluated under different interference conditions, causing phase-induced fluctuations to obscure the perturbation-induced response and making gradient estimation unreliable. This issue is particularly severe during the initial optimization stage when modal crosstalk is strong. Once the system reaches a low-crosstalk operating point, the influence of phase fluctuations is substantially reduced, enabling reliable gradient-based tracking. We therefore propose the two-stage optimization framework shown in Fig.~\ref{figure 3}a. In Stage I, a derivative-free genetic algorithm (GA) rapidly suppresses the initial modal crosstalk and drives the system into a low-crosstalk operating region. In Stage II, a gradient-descent-based algorithm continuously tracks environmental variations and maintains the optimal operating point, enabling real-time dynamic crosstalk compensation.

Stage I performs a derivative-free global search from a random initial state, where gradient estimation is unreliable. Although phase fluctuations perturb the measured kurtosis, they do not change its overall relationship with crosstalk: states with lower crosstalk still tend to exhibit smaller kurtosis. Therefore, kurtosis remains sufficiently informative to guide derivative-free global search. We thus employ the GA, which is well suited for noisy, high-dimensional, and nonconvex optimization ~\cite{jin2005evolutionary}. The population of N candidate heater-power vectors is randomly initialized. Each individual is evaluated by applying the corresponding voltages to the device and computing the kurtosis from the received signals. The population then evolves through selection, crossover, and mutation until a low-crosstalk operating point is reached. The detailed workflow is illustrated in Fig.~\ref{figure 3}a.


Stage II performs gradient-based real-time tracking starting from the low-crosstalk operating point reached in Stage I. In this regime, inter-channel interference is strongly suppressed, making the objective function less sensitive to phase fluctuations and enabling reliable gradient estimation. The heater parameters are updated sequentially, with one heater adjusted at each optimization step based on finite-difference gradient estimation. For the $i$-th heater, a small perturbation is applied to its heater power to estimate the gradient, followed by a gradient descent update:
\begin{equation}
    P_i \leftarrow P_i-\eta\nabla F(P_i),
\end{equation}
where $P_i$ denotes the power of the $i$-th heater and $\eta$ is the learning rate. Since the perturbation amplitude is sufficiently small, the induced signal variation remains negligible, allowing continuous crosstalk tracking without interrupting data transmission. Detailed information on the BSS algorithm is provided in APPENDIX~D.

\section{Experimental results}
\label{sec:experimental}

\subsection{Experimental Validation of the Two-Stage Algorithm}
\label{subsec:validation}

To validate the proposed two-stage algorithm, we conduct a real-time mode-demultiplexing experiment in which dynamic mode coupling is emulated by a MPC rotating at 0.2\(^\circ\)/s. In the initial computer-based validation, 100~Gbaud OOK signals are captured and digitally processed using a high-bandwidth ADC. Fig.~\ref{figure 3}c, d show the evolution of the cost function over the two optimization stages. In Stage~I, shown in Fig.~\ref{figure 3}c, the GA drives the system from its initial state to a low-crosstalk operating point within 34~epochs. The pronounced fluctuations in the convergence curve are mainly caused by phase-fluctuation-induced variations in the measured kurtosis. Despite these fluctuations, the GA still drives the system out of the high-crosstalk regime. As shown in Fig.~\ref{figure 3}e, both Channel~1 signal and Channel~2 signal change from closed eyes at the initial state to clearly open eyes after Stage~I convergence, and the optimized BERs decrease to \(3 \times 10^{-5}\) and \(2.1 \times 10^{-4}\) for the two channels, confirming substantial recovery of signal quality.

In Stage~II, as shown in Fig.~\ref{figure 3}d, the cost function remains at a consistently low level over 200~epochs, demonstrating robust long-term tracking and compensation. Notably, as shown in Fig.~\ref{figure 3}f, the eye diagrams remain open throughout a single GD epoch, during which the operating point \(P\), the perturbed point \(P + \Delta P\), and the updated point \(P - \eta \nabla F(P)\) are applied sequentially, with no noticeable degradation across the three power states.

These results show that the two-stage optimization enables effective adaptation during live data transmission. During Stage~I, the stochastic search rapidly drives the system toward a low-crosstalk operating region. After reaching this operating point, Stage~II provides stable gradient-based tracking while maintaining high-quality recovered signals throughout the adaptation process. In Stage~II, the heater updates are performed with small perturbations that introduce negligible signal degradation, allowing continuous tracking without interrupting data transmission. Experimental results therefore demonstrate real-time crosstalk tracking and compensation during live transmission.

\subsection{Real-Time Tracking and Compensation under Dynamic Perturbations}

\begin{figure*}[htbp]
\centering
  \includegraphics[width=1\linewidth]{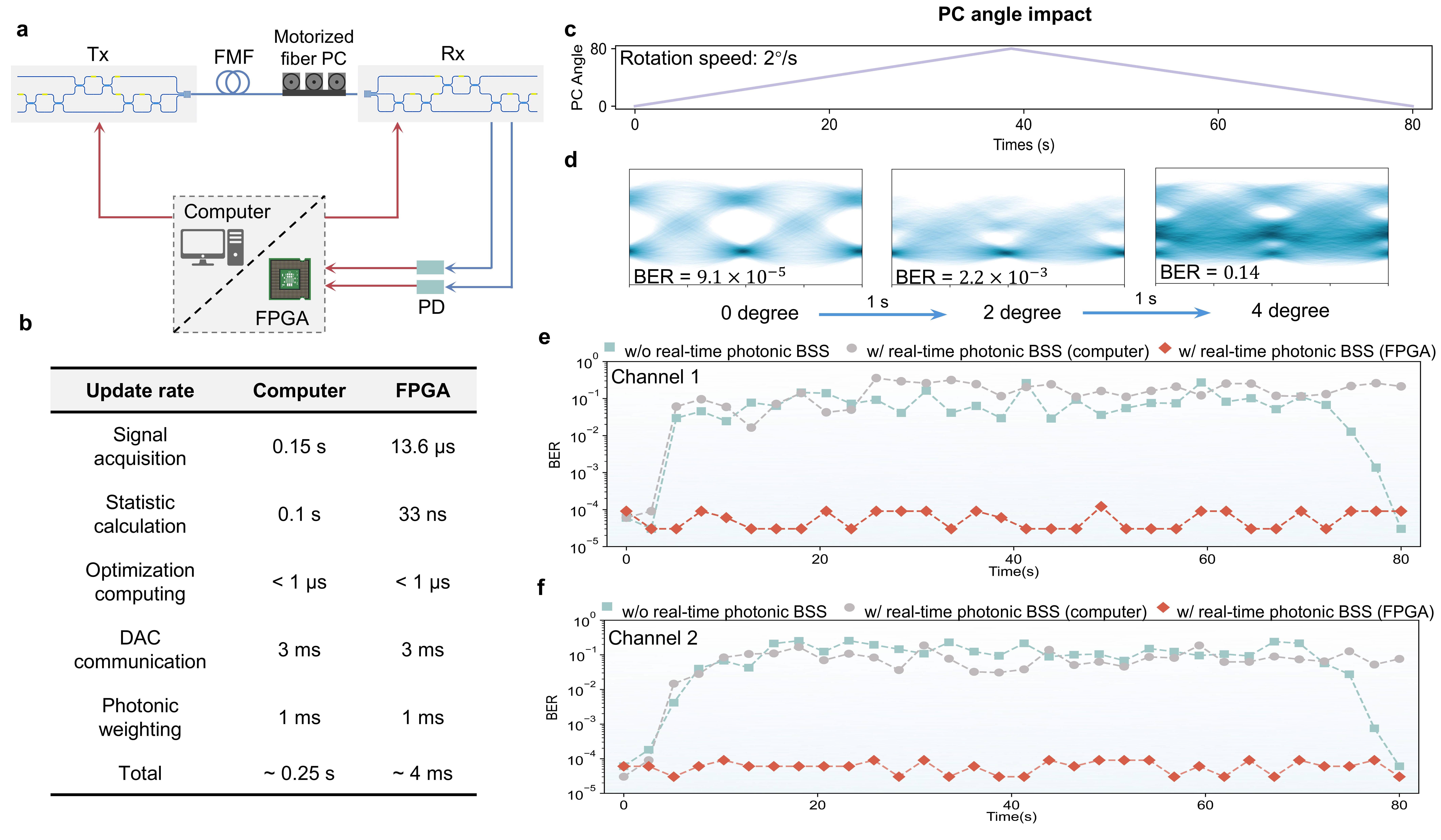}
  \caption{\textbf{FPGA-accelerated real-time photonic BSS for tracking dynamic channel perturbations.}
\textbf{a}, Experimental setup and control-latency comparison between computer- and FPGA-based implementations. The receiver outputs are detected by photodiodes and fed back to the controller for real-time update of the photonic processor. \textbf{b}, The table lists the per-iteration latency for signal acquisition, statistic calculation, DAC communication and photonic weighting. \textbf{c}, PC angle variation applied during experiment. \textbf{d}, Eye diagrams and BER measured at different PC angles. \textbf{e},\textbf{f}, 64 Gbaud OOK signals BER of the two channels versus time for three cases: without real-time photonic BSS, with computer-based real-time photonic BSS, and with FPGA-based real-time photonic BSS.
}
  \label{figure 4}
\end{figure*}

To evaluate the proposed method under faster dynamic perturbations and quantify the benefit of real-time FPGA processing, experiments are performed using both FPGA-based and computer-based implementations. The FPGA implementation details are described in APPENDIX~E. The 64 Gbaud OOK signals are acquired by a bandwidth-limited ADC backend with a 9~GHz analog bandwidth and a 2.4~GSa/s sampling rate, allowing evaluation of the proposed method under bandwidth- and sampling-rate-constrained conditions. A computer-based electrical backend is implemented as a benchmark to quantify the latency reduction enabled by FPGA-based processing. Fig.~\ref{figure 4} presents the FPGA-based experimental configuration, the latency comparison with a computer-based implementation, and the BER performance of different compensation schemes under dynamically varying modal crosstalk.

Environmental drift is emulated through controlled rotation of the MPC. We first characterize the impact of MPC rotation on signal quality before applying any compensation algorithm. As shown by the eye diagrams in Fig.~\ref{figure 4}d, even a 2$^\circ$ rotation degrades the BER from \(9.1 \times 10^{-5}\) to \(2.2 \times 10^{-3}\), while a 4$^\circ$ rotation increases it sharply to 0.14, with fully closed eyes.

In the dynamic crosstalk tracking and compensation experiment, the MPC continuously sweeps from 0$^\circ$ to 80$^\circ$ and back to 0$^\circ$ at 2$^\circ$/s over an 80-s period, corresponding to a ten-fold increase in perturbation speed compared with previous experiments in Section~5A. Fig.~\ref{figure 4}e and f show the measured BER evolution of both channels, with the corresponding MPC angle profile shown in Fig.~\ref{figure 4}c. Without photonic BSS compensation, the BER rapidly increases to approximately \(10^{-1}\), indicating complete link failure. The computer-based BSS cannot effectively track the dynamic drift, as its 0.25~s update interval is insufficient for tracking the channel variations, resulting in BER fluctuations between \(10^{-2}\) and \(10^{-1}\). In contrast, the FPGA-based BSS successfully tracks the perturbation and maintains BERs below \(10^{-4}\) for both channels throughout the entire perturbation cycle, enabling stable real-time crosstalk compensation under dynamic channel variations.

Importantly, the FPGA-based implementation achieves a substantial reduction in feedback-loop latency. Fig.~\ref{figure 4}b summarizes the latency breakdown of a single update iteration for the computer-based and FPGA-based implementations, including signal acquisition, statistic calculation, optimization computation, DAC communication, and photonic weighting. The total latency per iteration is reduced from 0.25~s in the computer-based implementation to 4~ms in the FPGA-based implementation, corresponding to a 62.5-fold speedup. This improvement is achieved by directly streaming the ADC output into the FPGA processing pipeline for real-time extraction of signal statistics, eliminating intermediate waveform buffering and computer-based post-processing. The largest latency reductions originate from the removal of software-based data transfer and computation overheads, with the signal acquisition and statistic calculation latency reduced from 0.15~s to 13.6~$\mu$s and from 0.1~s to 33~ns, respectively. These results confirm that the FPGA-based streaming architecture substantially reduces feedback-loop latency, enabling millisecond-scale adaptation of the photonic processor for dynamic crosstalk compensation.

\section{Discussion and Conclusion}

\subsection{Relaxed Electronic Requirements Enabled by Statistical Feedback}

\begin{figure}[!b]
\centering
  \includegraphics[width=0.5\linewidth]{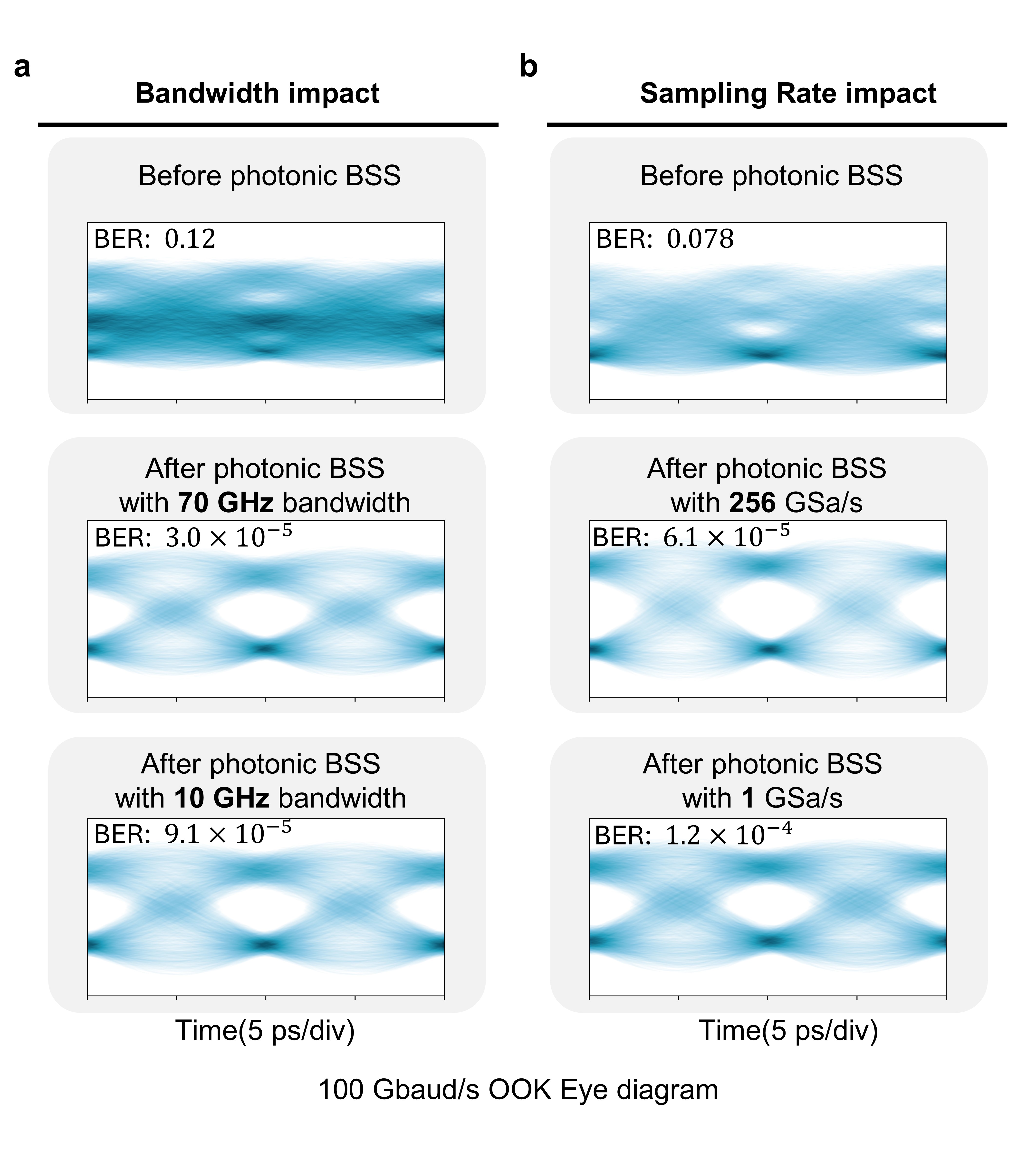}
  \caption{\textbf{Impact of bandwidth and sampling rate on photonic BSS performance.}
\textbf{a}, Eye diagrams before photonic BSS and after photonic BSS with receiver bandwidths of 70~GHz and 10~GHz. \textbf{b}, Eye diagrams before photonic BSS and after photonic BSS with sampling rates of 256~GSa/s and 1~GSa/s. All eye diagrams correspond to 100 Gbaud/s OOK signals.
}
  \label{figure 5}
\end{figure}

A notable practical advantage of the proposed photonic BSS is that its feedback loop can operate effectively with substantially relaxed bandwidth and sampling-rate requirements. This capability reflects a fundamental distinction from conventional DSP-based MIMO equalization, which relies on native baud-rate waveform capture to recover the temporal signal structure. To evaluate this experimentally, we intentionally apply digital filtering or resampling to the received signals within the experimental training loop before calculating the signal statistics. These operations allow us to evaluate reduced-bandwidth and reduced-sampling-rate conditions in the feedback path.

In the bandwidth study (Fig.~\ref{figure 5}a), the received 100 Gbaud OOK signals are digitally filtered to emulate electrical bandwidths progressively reduced from 70~GHz to 10~GHz, while the sampling rate is held constant at 256~GSa/s. After photonic BSS, the demultiplexed eye diagrams remain open for both electrical-bandwidth settings. The measured BER is \(3.0 \times 10^{-5}\) with a 70-GHz electrical bandwidth and \(9.1 \times 10^{-5}\) with a 10-GHz electrical bandwidth, confirming that the method remains effective even under bandwidth-limited detection. These results indicate that the feedback loop can sustain effective optimization under substantial bandwidth reduction. The required bandwidth is determined by whether the statistical features used for optimization can be reliably extracted after bandwidth reduction, rather than by the bandwidth required for full waveform reconstruction. A more detailed analysis of this bandwidth dependence is provided in the APPENDIX~F.

In the sampling-rate study (Fig.~\ref{figure 5}b), the sampling rate of the received signal is reduced from 256~GSa/s to 1~GSa/s, while the electrical bandwidth is kept 70~GHz. Even at 1~GSa/s, the demultiplexed eye diagram remains open after photonic BSS, with measured BERs of \(6.1 \times 10^{-5}\) at 256~GSa/s and \(1.2 \times 10^{-4}\) at 1~GSa/s. Theoretically, there is no strict lower bound on the sampling rate, because the method does not require full waveform information. However, for a given number of samples required to estimate the relevant statistics reliably, lowering the sampling rate increases the acquisition time and consequently lengthens the feedback-loop latency. Thus, in practice, the sampling rate should be chosen by balancing reduced ADC complexity against the latency budget of the adaptive feedback loop.

In addition to relaxing the ADC bandwidth and sampling-rate requirements, the FPGA implementation reduces the per-epoch update time to 4~ms. In the present setup, this latency is dominated by communication with the external DAC module, which accounts for about 3~ms per update. This communication time is mainly associated with the current external DAC interfacing scheme and could be further reduced in a more tightly integrated DAC--FPGA implementation ~\cite{altera2026agilex9}. In addition, the present 1~ms photonic weighting time is primarily determined by the need for thermal stabilization.With appropriate thermal design, this weighting time could be reduced to the tens-of-microseconds range. These considerations suggest that sub-millisecond adaptive operation is feasible in a more tightly integrated implementation. Such a reduction in update latency would increase the feedback bandwidth, allowing the system to track faster dynamic drift and maintain real-time crosstalk suppression under more rapidly varying channel conditions, such as the millisecond-to-microsecond crosstalk fluctuations typically observed in few-mode fibers ~\cite{chen2013characterization}.

\subsection{Scalability toward Higher-Order SDM Systems}

Although this work demonstrates real-time photonic BSS in a two-mode system, the framework is not fundamentally limited to two modes. The two-mode configuration is set by the available electronic interface---the current FPGA provides two ADC channels for statistical feedback---rather than by the photonic architecture or the BSS algorithm. For an $N$-mode system, an $N \times N$ universal MZI mesh implements the required unitary transformation in the optical domain, directly compensating the effective mode-coupling of the fiber regardless of its specific realization, while the electronic controller only extracts statistical features and updates the photonic weights.

Scaling to higher mode counts requires addressing several practical engineering challenges. A larger programmable MZI mesh is needed to realize higher-dimensional unitary transformations, which increases the number of phase shifters, control parameters, and accumulated optical loss. In addition, the enlarged optimization space may increase the complexity of parameter searching and feedback control. Nevertheless, these challenges arise from the implementation scale rather than from fundamental limitations of the proposed photonic BSS framework. The core advantage of the approach is preserved because the feedback loop relies on low-rate statistical features, with the update rate determined by the channel-variation timescale rather than the data symbol rate. Continued advances in large-scale silicon photonic integration, multi-channel electronic interfaces, and efficient optimization algorithms are expected to further support scaling toward higher-dimensional SDM systems.

A further consideration specific to few-mode fibers is modal degeneracy, where distinct modes share nearly identical propagation constants and the intra-fiber modal basis becomes non-unique. This does not limit the proposed approach: rather than recovering predefined physical modes, photonic BSS directly separates statistically independent data streams, so the exact modal basis need not be known, and any equivalent orthogonal transformation within a degenerate
subspace is automatically compensated as long as the transmitted channels remain independent.

Overall, this work demonstrates a photonic blind source separation system for adaptive modal crosstalk compensation in dynamic IM/DD space-division-multiplexed links, without relying on dedicated training sequences or full-waveform electronic acquisition. By exploiting intensity-domain statistical feedback and implementing the adaptive control loop on an FPGA, the proposed architecture substantially relaxes the ADC bandwidth and sampling-rate requirements while achieving millisecond-scale update latency. These capabilities provide a promising approach for addressing real-time crosstalk mitigation in future FMF communication systems. Compared with conventional DSP-based approaches, which require high-complexity waveform processing and face increasing power and latency challenges at ultra-high data rates, the proposed framework performs the critical linear transformation directly in the optical domain while avoiding the need for optical field reconstruction after direct detection. These results establish a practical pathway toward low-latency, hardware-efficient adaptive photonic processors for next-generation high-capacity optical interconnects.

\section*{APPENDIX}

\subsection*{A. Device Structure}
\label{sec:note1}

The photonic chips used in this work are fabricated on a silicon-on-insulator (SOI) platform through Advanced Micro Foundry (AMF), as shown in Fig.~2 of the main text. Each chip consists of a few-mode grating coupler, an MZI mesh, and single-mode grating couplers. The few-mode grating coupler provides vertical coupling between the spatial modes of the FMF and three on-chip waveguide channels, which are subsequently processed by the MZI mesh. The MZI mesh comprises three MZIs for programmable mode manipulation. The insertion loss of the few-mode grating coupler is approximately 10~dB, while each single-mode grating coupler has an insertion loss of approximately 4~dB. Since the overall system employs two chips, the total coupling loss is approximately 28~dB. The insertion loss of the current few-mode grating coupler can be further improved through additional design optimization. Previous demonstrations of optimized few-mode grating couplers have achieved insertion losses of approximately 4~dB, indicating the potential for reducing the optical interface loss and improving the overall system power budget~\cite{lu2024empowering}.

\subsection*{B. Few-mode Fiber}
\label{sec:note2}

The FMF used in this work is a commercial graded-index few-mode fiber designed for mode-division multiplexing transmission. The fiber operates in the C-band and supports the fundamental LP$_{01}$ mode and the higher-order LP$_{11}$ mode group. Under the weakly guiding approximation, the LP$_{11}$ mode group consists of two orthogonal spatial modes, LP$_{11a}$ and LP$_{11b}$, which are nearly degenerate and possess nearly identical propagation constants.

The key specifications of the FMF provided by the manufacturer are summarized in Table~S1. The fiber has a core diameter of 20~$\mu$m and a cladding diameter of 125~$\mu$m. The differential group delay (DGD) between the LP$_{01}$ and LP$_{11}$ modes is 0.14 ps/m. For the 5-m FMF link used in our experiment, the accumulated differential delay is negligible compared with the symbol duration of the transmitted signals. Therefore, the observed signal distortion is dominated by dynamic inter-modal crosstalk rather than modal dispersion.

\begin{table}[h] \centering \caption{Specifications of the commercial graded-index few-mode fiber.} \begin{tabular}{lc} \hline Parameter & Value \\ \hline Fiber type & Graded-index few-mode fiber \\ Operating wavelength & 1550 nm \\ Core diameter & 20 $\mu$m \\ Cladding diameter & 125 $\mu$m \\ Supported modes & LP$_{01}$, LP$_{11}$ mode group \\ Differential group delay & 0.14 ps/m \\ \hline \end{tabular} \end{table}

\subsection*{C. Design of the Nanoantenna-Array-Based Few-Mode Grating Coupler}
\label{sec:note3}

\begin{figure}[htbp]
\centering
\includegraphics[width=0.5\linewidth]{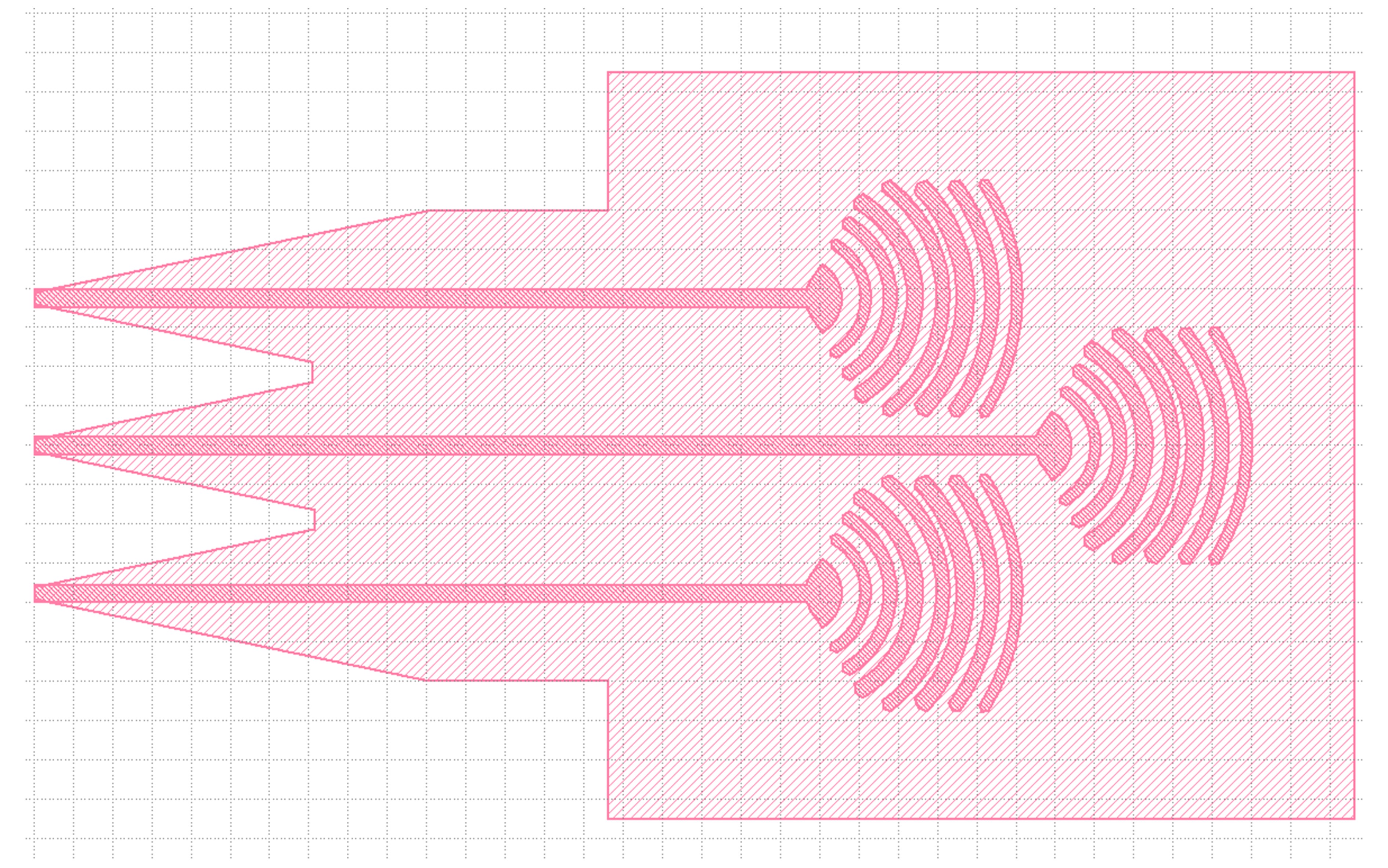}
\caption{
\textbf{Design of the nanoantenna-array-based few-mode grating coupler.}  GDS layout of the complete coupler formed on a
$19~\mu\mathrm{m}\times19~\mu\mathrm{m}$ silicon slab. The device consists of three identical sector-shaped nanoantenna arrays connected to three access waveguides.
}
\label{fig:s1}
\end{figure}

To interface the silicon photonic processor with the FMF, we designed a vertically coupled nanoantenna-array-based grating coupler for C-band operation, as shown in Fig.~\ref{fig:s1}. The coupler is formed on a $19~\mu\mathrm{m}\times19~\mu\mathrm{m}$ silicon slab and consists of three identical sector-shaped nanoantenna arrays connected to three access waveguides. The three nanoantenna arrays are positioned on a common circle centered at the center of the square silicon slab. Their spatial positions are therefore determined by the radius of this circle, denoted by $R_{\mathrm{pos}}$.

Each nanoantenna array contains $N$ concentric sector-shaped grating rows. Unlike a uniform grating coupler, the radial width and gap of each grating row are independently optimized. For the $i$th grating row, the radial tooth width and the gap to the subsequent row are denoted by $w_i$ and $g_i$, respectively. The corresponding local pitch $\Lambda_i$ is given by

\begin{equation}
\Lambda_i = w_i + g_i ,
\end{equation}

The device is simulated using three-dimensional finite-difference time-domain simulations. The optimization variables include the positioning radius $R_{\mathrm{pos}}$ and the radial tooth widths and gaps, $\{w_i\}$ and $\{g_i\}$, of a single nanoantenna array. The objective is to achieve balanced coupling efficiency among the three spatial modes supported by the FMF. 

For each input FMF spatial mode, the total coupling efficiency is calculated by summing the optical power collected by the three access waveguides:

\begin{equation}
\eta_m=\sum_{n=1}^{3}P_{m,n},
\end{equation}

where $P_{m,n}$ represents the coupled power from the $m$th FMF spatial mode to the $n$th access waveguide. To maximize the coupling performance of all spatial channels, a max-min optimization strategy is adopted, where the minimum coupling efficiency among the three modes is maximized:

\begin{equation}
\eta_{\mathrm{obj}}
=
\max
\left\{
\min(\eta_1,\eta_2,\eta_3)
\right\}.
\end{equation}

This optimization strategy prevents the design from preferentially improving only one spatial mode and ensures balanced coupling efficiency among all three FMF spatial channels.

\begin{table}[htbp]
\centering
\caption{Design parameters of the nanoantenna-array-based few-mode grating coupler.}
\label{tab:fmgc_design}
\begin{tabular}{lccc}
\hline
\multicolumn{4}{c}{\textbf{Global parameters}}\\
\hline
Parameter & Symbol & Value & Unit\\
\hline
Silicon device-layer thickness & $h_{\mathrm{Si}}$ & 220 & nm\\
Remaining slab thickness & $h_{\mathrm{slab}}$ & 90 & nm\\
Silicon slab dimensions & --- & $19\times19$ & $\mu$m$^2$\\
Number of nanoantenna arrays & --- & 3 & ---\\
Positioning radius & $R_{\mathrm{pos}}$ & 4 & $\mu$m\\
Number of grating rows per array & $N$ & 7 & ---\\
\hline
\multicolumn{4}{c}{\textbf{Nanoantenna row parameters}}\\
\hline
Row index & Tooth width $w_i$ & Gap $g_i$ & Period $\Lambda_i$\\
 & ($\mu$m) & ($\mu$m) & ($\mu$m)\\

\hline
1 & 0.262 & 0.475 & 0.737\\
2 & 0.324 & 0.335 & 0.659\\
3 & 0.412 & 0.258 & 0.670\\
4 & 0.332 & 0.341 & 0.673\\
5 & 0.444 & 0.182 & 0.626\\
6 & 0.346 & 0.288 & 0.634\\
7 & 0.260 & 0.335 & 0.595\\
\hline
\end{tabular}
\end{table}

The optimized design parameters of the nanoantenna-array-based few-mode grating coupler are summarized in Table~\ref{tab:fmgc_design}, including the global structural parameters, the row-dependent nanoantenna geometries.The mode-dependent coupling performance is evaluated using the mode-to-waveguide coupling matrix and the total coupling efficiency of each input spatial mode. The coupling matrix describes the distribution of the coupled optical power from each FMF spatial mode among the three access waveguides. The total coupling efficiency represents the fraction of the input optical power from each spatial mode collected by all three access waveguides, as summarized in Table~\ref{tab:fmgc_coupling}.

\begin{table}[htbp]
\centering
\caption{Simulated mode-dependent coupling performance of the nanoantenna-array-based few-mode grating coupler.}
\label{tab:fmgc_coupling}
\begin{tabular}{c|ccc|c}
\hline
 & WG$_1$ & WG$_2$ & WG$_3$ & Total coupling efficiency\\
\hline
LP$_{01}$ 
& 4.8\% & 6.5\% & 4.8\% & 16.1\% \\

LP$_{11a}$ 
& 5.4\% & 0.09\% & 5.4\% & 10.9\% \\

LP$_{11b}$ 
& 2.3\% & 7.8\% & 2.3\% & 12.4\% \\

\hline
\end{tabular}
\end{table}

\subsection*{D. BSS Algorithm}
\label{sec:note4}

The blind source separation (BSS) algorithm assesses the crosstalk condition using the statistical properties of the output signals. In this work, kurtosis is used as a metric to quantify the Gaussianity of the signal. The underlying principle is that the transmitted data streams are mutually independent and typically exhibit non-Gaussian distributions. After linear mode mixing, the received signals become more Gaussian due to the Central Limit Theorem. By monitoring the kurtosis of received signals, the photonic processor can be iteratively tuned toward the desired demixing state.

The kurtosis of a signal \(x\) is defined as
\begin{equation}
\mathrm{Kurt}(x)=\frac{\mathbb{E}\left[(x-\mu_x)^4\right]}{\sigma_x^4},
\end{equation}
where \(\mu_x\) and \(\sigma_x\) denote the mean and standard deviation of \(x\), respectively.

To improve the stability of the optimization, we further introduce a correlation-based term ~\cite{chen2023breaking}. Although kurtosis provides a useful indicator of the statistical state of each output, it does not by itself prevent two output channels from converging to the same source signal. Therefore, the correlation between different outputs is used to quantify their similarity. The correlation between two output signals \(y_i\) and \(y_j\) is defined as
\begin{equation}
\mathrm{Corr}(y_i,y_j)=\frac{\mathbb{E}\left[(y_i-\mu_i)(y_j-\mu_j)\right]}{\sigma_i \sigma_j},
\end{equation}
where \(\mu_i\), \(\mu_j\), \(\sigma_i\), and \(\sigma_j\) are the corresponding means and standard deviations.

In the implemented optimization, the objective function is defined as
\begin{equation}
\mathcal{F}=\left[\mathrm{Kurt}(y_1)+\mathrm{Kurt}(y_2)\right]\left[1+\mathrm{Corr}(y_1,y_2)\right].
\end{equation}
The optimization seeks to minimize \(\mathcal{F}\). In this formulation, minimizing the kurtosis term drives the output signals toward a more non-Gaussian distribution, while minimizing the correlation term suppresses solutions in which different outputs converge to the same source component. As a result, the combined objective improves the robustness and stability of the BSS process.

\subsection*{E. Real-time FPGA backend}
\label{sec:note5}

The real-time digital backend is implemented on a PCIe707 carrier board based on a Xilinx XCZU15EG Zynq UltraScale+ MPSoC, together with an FMC152B acquisition card equipped with a dual-channel 14-bit AD9689-2600 ADC operated at \(2.4~\mathrm{GSa/s}\). The acquisition front end streams digitized data to the FPGA through an 8-lane JESD204B Subclass 1 interface, after which all high-rate reception, stream formatting, and statistical processing are performed entirely within the programmable logic (PL). Reduced statistical metrics are exported to the processing system (PS) through an AXI-Lite register interface, whereas the PS is restricted to supervisory readback, parameter updates, and outer-loop optimization. This partitioning separates the architecture into a high-throughput PL datapath and a low-rate PS control plane, thereby keeping the primary sample stream inside the FPGA fabric without an off-chip memory round-trip.

The core computation is organized as a streaming statistical reduction engine rather than a memory-centric frame processor. At each clock cycle, the engine accepts two 8-lane groups of 14-bit samples and continuously updates alignment buffers together with kurtosis and cross-correlation accumulators. At the end of each observation window, the PL enters a frame-boundary reduction stage in which the accumulated moments are normalized and converted into compact correlation metrics. Continuous moment generation is realized using parallel fixed-point multipliers that independently compute the fourth powers and cross-correlations for the 8 input lanes. To sustain high-throughput operation, the arithmetic datapath is explicitly pipelined, resulting in a deterministic latency of exactly 10 clock cycles. The input samples are first squared using an initial multiplier with a 3-cycle pipeline delay, and the resulting squared values are then processed by a second 6-cycle pipelined multiplier. These parallel multiplier stages therefore account for 9 clock cycles of the total latency. A fully combinational 3-stage binary adder tree spatially reduces the 8 parallel outputs into a single scalar sum without adding extra cycle latency. The resulting sum is subsequently accumulated by a synchronous register-based accumulator controlled by a periodic reset signal. By accumulating the 8 parallel lanes over the full observation window, the engine processes exactly 32{,}768 discrete samples. The subsequent mean normalization is therefore implemented as a combinational 15-bit arithmetic right shift, avoiding the latency and DSP overhead of a hardware divider. Finally, the normalized metric is captured by a terminal output register, which contributes the 10th and final cycle of the complete datapath latency. With a PL clock frequency of \(300~\mathrm{MHz}\), the 10-cycle pipeline depth corresponds to an absolute hardware delay of approximately \(33.3~\mathrm{ns}\). For a 4096-cycle observation window, each statistical update spans approximately \(13.6~\mu\mathrm{s}\). Including the complete reduction tail, the overall frame-to-metric latency is approximately \(13.63~\mu\mathrm{s}\). The implemented datapath thus combines on-the-fly accumulation with frame-boundary metric emission while fully avoiding frame materialization in external memory.

\begin{figure}[htbp]
    \centering
    \includegraphics[width=0.9\textwidth]{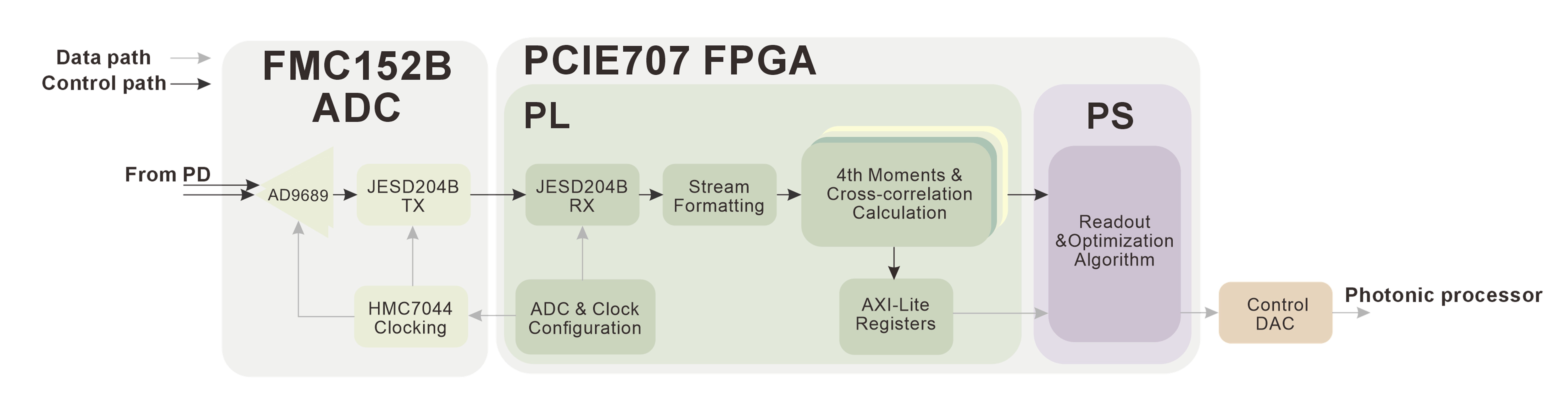}
    \caption{{FPGA-based real-time digital processing architecture.}}
    \label{fig:s2}
\end{figure}

\subsection*{F. Kurtosis Dependence on Bandwidth and Sampling Conditions}
\label{sec:note6}

\begin{figure}[htbp]
    \centering
    \includegraphics[width=\textwidth]{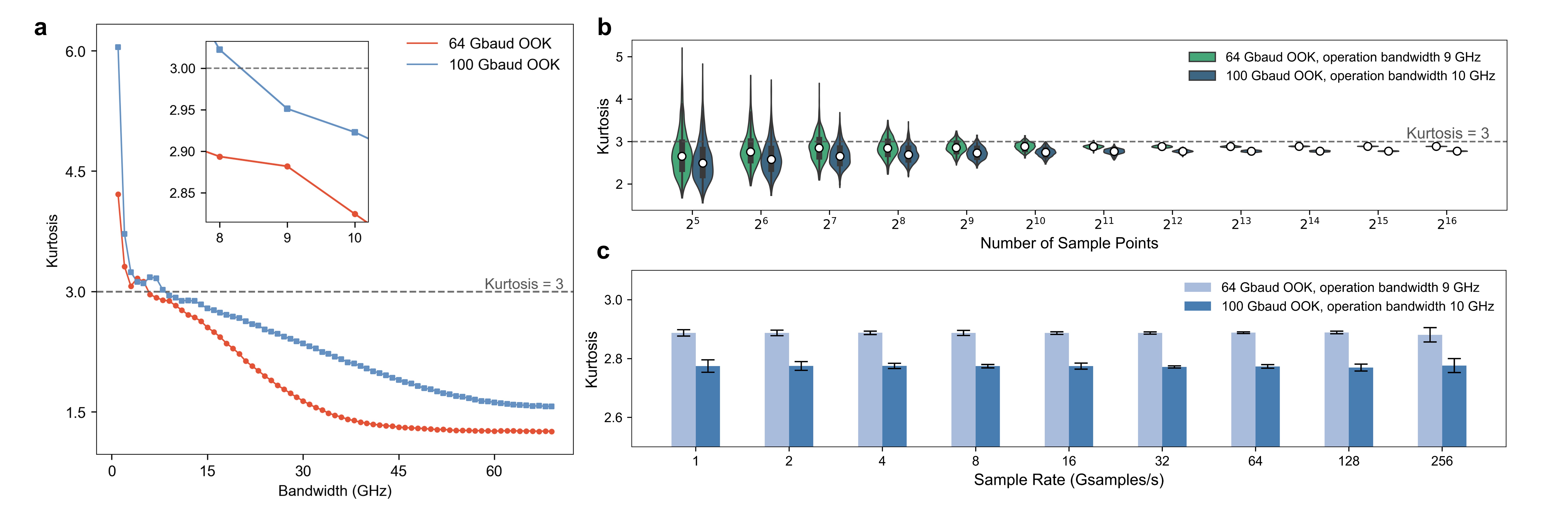}
    \caption{{Kurtosis characteristics of 64~Gbaud and 100~Gbaud OOK signals under different operation bandwidth and sampling conditions.} 
    (a) Kurtosis as a function of operation bandwidth. 
    (b) Dependence of the kurtosis estimate on the number of sample points. 
    (c) Dependence of the kurtosis estimate on sampling rate.}
    \label{fig:s3}
\end{figure}

Fig.~\ref{fig:s3}a summarizes the dependence of the kurtosis metric on the receiver operation bandwidth for the 64~Gbaud OOK and 100~Gbaud OOK signals used in this work. As shown in Fig.~S3a, the kurtosis of both signals varies systematically with the receiver bandwidth due to bandwidth-induced waveform distortion. The dashed line indicates the Gaussian reference value of 3. As the receiver bandwidth is reduced, the signal distribution becomes increasingly Gaussian, causing the kurtosis to approach the Gaussian reference value and reducing the statistical contrast required for reliable BSS optimization. In contrast, when the original signal maintains a sufficient deviation from the Gaussian reference, the statistical distinction remains robust even after bandwidth limitation and crosstalk-induced mixing.

Since the ADC used in the FPGA-based feedback platform provides an analog bandwidth of 9~GHz, the kurtosis of both baud-rate signals was evaluated under this hardware constraint. At 9~GHz bandwidth, the kurtosis of the 100~Gbaud OOK signal becomes close to the Gaussian reference value, indicating insufficient non-Gaussianity for reliable optimization. In comparison, the 64~Gbaud OOK signal maintains a deviation greater than 0.1 from the Gaussian reference, providing sufficient statistical margin for BSS-based feedback. Therefore, the real-time FPGA demonstration was performed using 64~Gbaud OOK signals rather than 100~Gbaud OOK signals.

The effect of the number of sample points used for kurtosis estimation is shown in Fig.~\ref{fig:s3}b. The results show that the estimated kurtosis gradually converges as the observation length increases and becomes sufficiently stable at \(2^{15}\) sample points. This value was therefore adopted in both the experiments and the FPGA real-time implementation. The dependence of kurtosis on sampling rate is shown in Fig.~\ref{fig:s3}c. The results indicate that the kurtosis estimate is largely insensitive to the sampling rate over the investigated range, demonstrating that the metric is robust with respect to temporal sampling density. Accordingly, the sampling rate can be selected primarily based on hardware convenience rather than estimation accuracy. In the present experiments, a sampling rate of \(2.4~\mathrm{GSa/s}\) was used.

\begin{backmatter}
\bmsection{Funding}
This work was supported by RGC YCRG C4004-24Y, C1002-22Y, ECS 24203724, GRF 14208925, STG 3/E-404/24-N, NSFC 62405258, ITF ITS/237/22, National Key Research and Development Program of China 2024YFE0203600, NSFC/RGC N\_CUHK444/22. This work was also funded by Hong Kong  ITC  under  Raise+ project number RAI/23/1/028A.

\bmsection{Disclosures}
The authors declare no conflicts of interest.

\bmsection{Data availability} Data underlying the results presented in this paper are not publicly available at this time but may be obtained from the authors upon reasonable request.


\end{backmatter}

\bibliography{sample}

\end{document}